\documentclass[pra,superscriptaddress,,nofootinbib]{revtex4}
\usepackage{blindtext}
\usepackage{amssymb}

\usepackage{amsmath}
\usepackage{graphicx}
\graphicspath{ {images/} }

\newtheorem{prop}{Proposition}\def\PRO{\begin{prop}}\def\ORP{\end{prop}}
\newtheorem{coro}{Corollary}\def\COR{\begin{coro}}\def\ROC{\end{coro}}
\newtheorem{theo}{Theorem}\def\TH{\begin{theo}}\def\HT{\end{theo}}
\def\TH{\begin{theo}}\def\HT{\end{theo}}
\newtheorem{Definition}[prop]{Definition}\def\DE{\begin{defi}}\def\ED{\end{defi}}

\newtheorem{lemme}[prop]{Lemma}\def\LE{\begin{lemme}}\def\EL{\end{lemme}}

\begin{document}
%\doinum{10.3390/e17085635}
\title{Unconditionally Secure Quantum Signatures}
\author{Ryan Amiri}
\email{ra2@hw.ac.uk}
\author{Erika Andersson}
\affiliation{SUPA, Institute of Photonics and Quantum Sciences, Heriot-Watt University, Edinburgh EH14 4AS, United Kingdom}

\begin{abstract}
Signature schemes, proposed in 1976 by Diffie and Hellman, have become ubiquitous across modern communications.~They allow for the exchange of messages from one sender to multiple recipients, with the guarantees that messages cannot be forged or tampered with and that messages also can be forwarded from one recipient to another without compromising their validity. Signatures are different from, but no less important than encryption, which ensures the privacy of a message.~Commonly used signature protocols---signatures based on the Rivest--Adleman--Shamir (RSA) algorithm, the digital signature algorithm (DSA), and the elliptic curve digital signature algorithm (ECDSA)---are only computationally secure, similar to public key encryption methods. In fact, since these rely on the difficulty of finding discrete logarithms or factoring large primes, it is known that they will become completely insecure with the emergence of quantum computers.~We may therefore see a shift towards signature protocols that will remain secure even in a post-quantum world.~Ideally, such schemes would provide unconditional or information-theoretic security. In this paper, we aim to provide an accessible and comprehensive review of existing unconditionally secure signature schemes for signing classical messages, with a focus on unconditionally secure quantum signature~schemes.
\end{abstract}
\maketitle

\section{Introduction}

Non-orthogonal quantum states cannot be perfectly distinguished from each other, nor perfectly copied, and the precision of both preparation and measurement of quantum states is limited by uncertainty relations.~Such counter-intuitive properties might at first seem to pose limits on practical applications, but can in fact also be useful.~The security of quantum cryptography rests exactly on such quantum-mechanical features.~Arguably, research on quantum cryptography started with Wiesner's work on ``unforgeable subway tokens'' \cite{Wiesner} (which was actually only published long after its conception; see \cite{Brassardtalk} for a short, but entertaining account of early research on quantum cryptography). After Bennett's and Brassard's 1984 paper on quantum key distribution (QKD)~\cite{BennettBrassard}, quantum cryptography began to receive more interest, and today, QKD is one of the most active and furthest developed topics in quantum information science~\cite{QKDreview, QKDreview2}. 

Modern cryptography, however, encompasses much more than encryption of messages~\cite{PaarBook}. Functionalities such as authentication, signatures, oblivious transfer, bit commitment and Byzantine agreement are important, for example, in communication, for multi-party computation and for secure voting schemes.~Early work on quantum cryptography also looked to applications other than encryption~\cite{Brassardtalk}, but
nevertheless, the term ``quantum cryptography'' is often used as synonymous with quantum key distribution.~In recent years, however, applications other than QKD have again begun to receive more interest, both theoretically and experimentally.~The focus of this review will be quantum signature schemes.~Signature schemes, proposed in 1976 by Diffie and Hellman~\cite{HellmanDiffie}, are crucial to digital communications and have become ubiquitous in the modern world.~Their aim is to provide a way to securely sign (classical) messages, so that they cannot be forged or tampered with.~Crucially, in a signature scheme, messages are also transferable, meaning that a recipient of a signed message can check whether another recipient is likely to accept that message if it is forwarded, without there and then contacting the other recipient. Broadly speaking, this distinguishes signature schemes from authentication schemes, which ensures two communicating parties that messages have not been tampered with, but without necessarily guaranteeing transferability. Since we are interested in signing classical messages, this review will not cover schemes designed to authenticate or sign quantum messages, such as \cite{BarnumAuth,LuFengSig}. We also do not cover either classical or quantum schemes for so-called blind signatures, where a signer does not learn about the message being signed~\cite{BlindSig}.

Two-party authentication of classical messages can be efficiently accomplished ``classically'' with information-theoretic security \cite{WegmanCarter}; this is indeed used also in full implementations of QKD. Commonly used digital 
signature schemes, however, only provide computational security, {relying on public key cryptography}.~While security is only computational, it implies ease of use, since public keys can be easily distributed, for example, by a certificate authority.~Quantum signature schemes for signing classical messages, on the other hand, can be made information-theoretically secure, similar to the information-theoretic security of QKD. ``Information-theoretic'' or ``unconditional'' security means that security does not rest on computational assumptions, but instead can be proven to hold; in the case of QKD and quantum signatures, security is guaranteed by the laws of quantum mechanics. These protocols remain secure as long as adversaries are bound by what is possible according to quantum mechanics. Naturally, it is important that the implementation of such protocols does what it is supposed to, otherwise loopholes may arise.

It is also possible to construct ``classical'' unconditionally secure signature (USS) schemes~\cite{ChaumUSS, HanaokaSig, DunjkoQKDComp}. We thus have to ask what the advantages of {quantum} unconditionally secure schemes might be, given that implementing a quantum protocol is usually more cumbersome than implementing a classical protocol. The answer to this question is not fully known, but the advantage of 
quantum signature schemes may lie, e.g., in what resources are needed or in exactly which parties have to be connected by what type of communication channels and in how many of, e.g., pairwise quantum channels are needed. All known information-theoretically secure ``classical'' signature schemes also use shared secret keys, which would have to be obtained in an information-theoretically secure way, for example using QKD, making them indirectly rely on quantum features.~In addition, the scheme in \cite{ChaumUSS} uses an authenticated broadcast channel, and the scheme in \cite{HanaokaSig} is phrased in terms of a trusted omnipotent initialiser who sets up the scheme and who would have unlimited power to forge messages, aid in repudiation, and so on. Quantum signature schemes may be of interest mainly if one wants to do without such extra resources and trust assumptions. For this reason, in this review we will not cover, for example, quantum signature schemes that make use of a trusted third party~\cite{ThirdPartyQuantum}, since if this exists, classical unconditionally secure signature schemes are possible.
Some confusion in terminology exists {as to exactly what defines the functionality of a signature scheme or digital signature scheme. One may also take the view that a ``signature'' or ``digital signature'' must have the public verifiability implied by public key cryptography and that ``unconditionally secure signatures'' should instead be called perhaps ``transferable message authentication codes'', or similar, but we have not found this choice of terminology in the literature. Instead, we are following existing terminology introduced in \cite{ChaumUSS, GottesmanQDS, SwansonStinson}.}

The aim of this review is to provide a short and accessible account of quantum signature schemes, mainly intended for readers with a background in quantum physics.~To date, there is very little work done on unconditionally secure signature schemes, whether ``classical'' or quantum, and with this review, we hope to inspire more work on this topic.~We start by defining basic properties of signature schemes and outlining how public key-based (``classical'') signatures work.~We then informally explain what one-way and hash functions are and how these can be used in signature schemes, followed by some remarks on existing unconditionally secure ``classical'' signature schemes. We then proceed to talk about quantum signature schemes, starting with the initial work by Gottesman and Chuang~\cite{GottesmanQDS} and followed by subsequent work on quantum signature schemes that are more suitable for real applications. {Both experimental and theoretical work is reviewed. Far from all proposed schemes have been realized; we mention all experimental full or partial realisations of quantum signatures of which we are aware.}
At the end, we %mention efforts to construct reusable quantum public key schemes, and 
briefly discuss open problems and future research directions.

\section{Basic Properties of Signature Schemes}
\label{sec:basic}

Signature protocols have three aims: message integrity (the message has has not been altered in transit); message authentication (that the sender of the message is authentic); and non-repudiation (the sender cannot deny the creation of a message) \cite{PaarBook}. Depending on the mechanism for dispute resolution, non-repudiation is related to transferability, which means that a recipient of a message can check whether it is likely to be accepted by another recipient if the message is forwarded.~{Informally}, a signature scheme is secure if it has the following three properties \cite{SwansonStinson}:
\begin{enumerate}
\item [(1)] Unforgeability: A dishonest party should not be able to successfully send a message pretending to be someone else.
\item [(2)] Non-repudiation: A signer should be unable to successfully deny that he sent a message signed with his signature.
\item [(3)] Transferability: If a verifier accepts a signature, he should be confident that any other verifier (e.g., a judge) would also accept the signature.
\end{enumerate}

It is important to stress that for transferability, a recipient should be able to test, upon receipt of a signed message and without any further interaction with any other party, whether one or more other recipients are likely to accept the message if it is forwarded.~{Formal security definitions do not currently exist for general quantum signature protocols. Instead, authors have used the above criteria to show that, for their specific protocol, participants are unable to forge or repudiate and that messages are transferable, except with negligible probability.~Recently, Swanson and Stinson \cite{SwansonStinson} produced a formal description of the security requirements of ``classical'' unconditionally secure signature schemes (described in Section~\ref{sec:classical}).~These definitions can be adapted to quantum signature schemes \cite{JuanMiQCrypt} (described in Section \ref{sec:quantum}), albeit with some modifications.~For example, in \cite{SwansonStinson}, it is implicitly assumed that there is only one signature for each message that will pass verification with all users, whereas for existing quantum signature schemes, there may be more than one such signature, without altering the actual practical functionality of the scheme.}

Since messages may be transferred from one recipient to another, in the minimal signature scenario, classical or quantum, there are three parties: one sender, whom we will refer to as ``Alice'', and two recipients, referred to as ``Bob'' and ``Charlie''.~It is important to stress that, unlike in QKD, where sender and receiver are assumed to be honest, in a signature protocol, any party could be dishonest.~This means that there must be a pre-agreed dispute resolution procedure.~For example, imagine a three-party scenario where Alice is dishonest and sends a valid signature to Bob who accepts it.~At some point in the future, Alice tries to repudiate and claims that she did not send the signed message.~If Bob maintains that she did, how do they decide who is lying?~An obvious solution, and the one assumed for the quantum signature protocols presented in this review, is for all participants to vote and take the majority decision. In this example, then, Bob would forward the signed message on to Charlie, who would accept it as valid (by transferability).~Then, both Bob and Charlie would correctly decide that Alice did in fact send the message. Majority voting should, if the scheme is correctly designed, lead to the correct outcome whenever at least half of the participants are honest.~It can be seen that in the three-party case with majority voting, non-repudiation and transferability are closely linked.~Any such three-party scheme satisfying unforgeability and transferability will automatically be secure against repudiation attempts. {Therefore, when proving security in this scenario, one must show that for the protocol considered, two conditions hold. First, the probability of Bob (or Charlie) being able to find a message-signature pair (which has not come from Alice) that Charlie (or Bob) will accept as valid is negligibly small. This usually means showing that the probability decays exponentially in the size of the signature length. Second, one must show that if Bob (or Charlie) accepts a message-signature pair as valid, except with negligible probability, Charlie (or Bob) will also accept the message-signature pair as valid. As mentioned above, general security definitions are a recent development, and as such, existing security proofs have not been written in terms of the new security framework.}

In general, there are many possible dispute resolution procedures, each with different trust assumptions.~One may have a single arbiter (a trusted party) whose decision is final in resolving disputes. Alternatively, one may decide that certain parties are more trustworthy than others and implement a weighted voting system. The choice of dispute resolution method will depend on the specific scenario in which the signature scheme will be used. It is expected that dispute resolution will be used relatively rarely and only as a last resort. If majority voting is used, all participants must be contacted, which may take considerable effort. If a trusted party resolves disputes, dispute resolution is akin to going to court.

\section{``Classical'' Signature Schemes} \label{sec:classical}

\subsection{Public-Key Digital Signatures}

A widely used concept in classical cryptography is the one-way function.~Informally, this is a function whose output is easy to compute given an input, but whose input (pre-image) is hard to compute given an output.~A trapdoor function is a one-way function whose pre-image becomes easy to compute given an output, as well as an associated secret.~A standard example is prime factorisation: given two large prime numbers, it is easy to compute their product, but given their product, it is believed to be computationally difficult to find the prime factors, unless more information is given as well. This is the basis of Rivest--Adleman--Shamir (RSA)~\cite{PaarBook} encryption schemes, where, roughly speaking, the product of the two primes is the public key, and the prime numbers themselves are the private key. Anyone with access to the public key can encrypt messages, but only those with access to the private key can decrypt the messages.~No one-way function has been proven to be information-theoretically secure. In fact, proving the existence of such a one-way function would imply $P\neq NP$.

Roughly speaking, then, if a message can be correctly decrypted using the public key belonging to a particular sender, then it follows that the message must have been encrypted using the private key for that sender;~hence that the message originated from this sender and has not been altered.~Any honest party in possession of the correct public key will reach the same conclusion, implying that messages are transferable.~In practice, it is common to create a signature scheme using slightly modified versions of existing public key encryption methods.~The most commonly used digital signature schemes are the digital signature algorithm (DSA) \cite{DSA} and the elliptic curve digital signature algorithm (ECDSA)~\cite{ECDSA}, which have been the standard in the U.S. since 1998. Although slightly different from RSA encryption, the general principle is the same, with security derived from the assumed computational difficulty of finding discrete logarithms.~Given the widespread usage of RSA, DSA and ECDSA, it was perhaps unsettling (and exciting) when it was shown \cite{ShorFactor} that quantum computers could efficiently factorise prime numbers and find discrete logarithms. This means that any scheme relying on these problems will become completely insecure in a post-quantum world. 

A natural question is then: are there any signature protocols, classical or quantum, that are secure if quantum computers exist?~One solution is to start using quantum signature schemes to generate unconditional security, but there may also be entirely ``classical'' encryption and signature methods that remain computationally secure in the presence of quantum computers~\cite{Bernstein}.~Even if quantum computers do not yet exist, perhaps we should already start worrying.~In order to protect against future security breaches by an eavesdropper listening in on communications now and breaking the cryptographic scheme later on when better algorithms or quantum computers become available, perhaps we should already be using more secure cryptographic protocols.~This can be used as an argument for quantum key distribution, with the aim of using the generated secret key for message encryption using the one-time pad, which essentially is the only encryption method that is provably information-theoretically secure~\cite{onetimepad}. For signature schemes, the issue with future security breaches is still present, but somewhat less serious. If we anticipate that a signature scheme is about to be broken, we can re-sign the data using a stronger signature scheme. However, if a signature scheme has already been broken, then 
security is lost; an adversary may have tampered with the data and forged a signature, and obviously, re-signing with a stronger scheme no longer helps.

\subsection{Cryptographic and Universal Hash Functions}

One-way functions are closely related to hash functions, and both can be used for signature schemes and authentication. In modern computer science, there are several types of hash functions with different properties depending on their intended usage. 
Roughly speaking, a hash function is an easily computable function that maps a longer message to a shorter string, called a hash or a tag, of shorter, often fixed, length.
Given a tag, it may or may not be easy to compute a message corresponding to the tag; that is, the hash function may or may not be a one-way function. A simple example where it is easy to find a message corresponding to a given tag is if the tag simply is the XOR of all of the message bits. Strictly speaking, such a checksum or hash sum is again slightly different from a hash function in its detailed properties, but this example nevertheless serves well as an illustration. Checksums are useful, e.g., to detect unintentional mistakes in a message, such as a bank account number or social security number. Another application where it is not necessary that a hash function is hard to invert is if it is used to accelerate data lookup. A fingerprint is another related concept; this maps a large data item to a much shorter string, to be used, for example, to facilitate a comparison between large data items.

For authentication and signature schemes, hash functions where it is difficult to find messages corresponding to a given tag are of more interest.~There are two main types of such hash functions, cryptographic hash functions and universal hash functions.~These are special types of one-way functions commonly used in computer science and are particularly useful in many cryptographic protocols.~As explained below, cryptographic hash functions can be used for signature schemes, just as any one-way function can.~Two important situations where universal hash functions can be used are message authentication with information-theoretic security \cite{WegmanCarter} and privacy amplification in QKD \cite{RennerPrivamp}.

Somewhat more formally, we can demand that a cryptographic hash function should have the following three properties \cite{PaarBook}:
\begin{enumerate}
\item[(1)] Pre-image resistance: Given $h(x)$, it should be %computationally 
difficult to find $x$, that is, these hash functions are one-way functions.
\item[(2)] Second pre-image resistance: Given $x_1$, it should be %computationally 
difficult to find an $x_2$, such that $h(x_1) = h(x_2)$.
\item[(3)] Collision resistance: It should be %computationally 
difficult to find any distinct pair $x_1, x_2$, such that $h(x_1)=h(x_2)$.
\end{enumerate}

Since there are no one-way functions that are known to be provably 
more difficult to invert than to compute, the security of cryptographic hash functions is computational.~This also then holds for signature and authentication schemes constructed from cryptographic hash functions.

Somewhat different from a cryptographic hash function, a universal hash function is instead a collection of hash functions, from which one picks one particular function. Exactly which hash function one picks is determined by a secret shared key. Individually, the hash functions in the set do not have to satisfy Properties 1--3 above. Given knowledge of a certain number of messages and their hash values, but without knowledge of the specific hash function chosen, the hash values behave like independent random variables. That is, knowledge of the hash of a certain number of messages gives no information regarding the hash value of any other distinct message. To come up with another message-hash pair, one can do no better than guessing (or if the function is ``almost universal'', at least not much better than this). 

More formally, suppose $H$ is a set of hash functions, mapping longer messages to shorter hashes. $H$ is strongly universal$_n$ if given any $n$ distinct messages $x_1, x_2, \ldots , x_n$ and (not necessarily distinct) hashes $y_1, y_2, \ldots y_n$, then the number of hash functions taking $x_1$ to $y_1$, $x_2$ to $y_2$, \emph{etc.},~is equal to $|H|/|T|^n$, where $|H|$ is the number of hash functions and $|T|$ the number of possible hashes.~It is this property of universal hash function sets that makes them useful in protocols desiring information-theoretic security.~Wegman and Carter~\cite{WegmanCarter} give a construction by which it is efficient to specify which of the hash functions in the set one picks. If the messages have $n$ bits and hashes have $t$ bits, then this takes only of the order $t\log n$ bits. The scheme is ``almost strongly universal'', and Wegman and Carter describe how this can be used, e.g.,~to construct an efficient unconditionally secure authentication scheme. It is this type of authentication scheme that is used in a full implementation of QKD, in order to make the key generation (or key expansion) unconditionally secure.

\subsection{Lamport--Diffie One-Time Signatures}

An interesting class of digital signatures, closely related to the proposed schemes for quantum digital signatures, are hash-based digital signatures.~Lamport \cite{LamportSig} introduced the concept of a one-time signature scheme that can be securely implemented using any collision-resistant one-way function. Collision-resistance was defined above and means, loosely speaking, that the probability for two (or more) input values is sufficiently unlikely to be mapped to the same function value. To illustrate such a scheme, imagine that Alice wants to send a single signed bit, zero or one, at some point in the future. She will choose two random inputs, $k_0$ and $k_1$, to a collision resistant 
one-way function $f$ and compute $f(k_0)$ and $f(k_1)$. The public key is then $\{(0, f(k_0)), (1, f(k_1))\}$. Since the function is assumed to be one-way, potential forgers cannot find an input generating $f(k_0)$ or $f(k_1)$. To send a signed one-bit message, $b$, Alice would send $(b, k_b)$. The recipient would apply the publicly known $f$ to $k_b$ and accept the message only if $f(k_b)$ matches the public key. Once the message is sent, the public key cannot be re-used and must be discarded, hence the name ``one-time signature scheme''. 

In \cite{Merkle}, Merkle extended the one-time signature scheme to make it re-usable, though only for a finite number of messages. This inefficiency in terms of public key reusability meant that hash-based digital signatures have been largely ignored in favour of the more efficient ECDSA scheme. However, since ECDSA will be insecure in a post-quantum world, hash-based signatures are gaining in popularity as they seem to be much more resistant to attacks from quantum computers (we note here that there are alternatives to hash-based signatures in post-quantum cryptography, for example lattice-based cryptography, but these similarly provide only computational security and are not considered in this review). Still, to prove the security of these post-quantum schemes, it must be {assumed} that the function used is a (classical) one-way function resistant to efficient quantum inversion algorithms \cite{SongPQ}.~Again, it should be stressed that it has never been proven that such a function exists, even for classical computers; it is only an expectation. Nevertheless, efficient quantum algorithms are difficult to find, and although there is an efficient quantum algorithm for factoring, making RSA insecure, it is hoped that hash-based signatures will remain computationally secure, even in a post-quantum world.

Even assuming the existence of one-way functions, all such post-quantum signature schemes provide only computational security and can be broken with enough time or computational power. Quantum signatures and classical unconditionally secure signature schemes, on the other hand, provide unconditional security.~The first version of quantum signature schemes, including the one in~\cite{GottesmanQDS}, are in fact quantum analogues of the Lamport--Diffie one-time signature scheme.~They take advantage of the fact that quantum mechanics can give us provably secure one-way functions, as explained below in Section \ref{Sec:oneway}.

\subsection{unconditionally Secure ``Classical'' Signature Schemes} \label{sec:USCS}

All widely used classical digital signature schemes give only computational security, but there is no fundamental reason that this should be the case. In fact, completely classical schemes for unconditionally secure signatures (USS) have been proposed, although this is by no means a well-investigated research topic.~This is mainly due to the popularity of public key-based signature schemes, due to their ease of use and favourable scaling properties.~As a consequence, there are still no widely accepted precise definitions of what it means for a protocol to be an unconditionally secure signature scheme and exactly what security requirements must be satisfied.~In this review, we follow the definitions and security requirements set out in \cite{SwansonStinson}, which were stated informally in Section \ref{sec:basic}.~Since quantum signature schemes very likely are more cumbersome to implement than any classical scheme, ideally to motivate us to investigate quantum signature schemes, these should have some advantages over classical unconditionally secure signature schemes.~It is therefore important to know exactly what the properties of classical USS schemes~are.

Chaum and Roijakkers were the first to propose an USS scheme~\cite{ChaumUSS}.~Given an authenticated (classical) broadcast channel and pairwise secret authenticated (classical) channels, this scheme makes use of the untraceable sending protocol from \cite{ChaumDining} in order to send a single signed bit.~In order to send longer messages, the protocol should be iterated, leading to a signature length that scales linearly with the size of the message.~This protocol can be made unconditionally secure, because it does not rely on the use of assumed one-way functions.~Instead, the sender's signature contains elements from all participants sent anonymously using the untraceable sending protocol.~Intuitively, this is what guarantees transferability: a dishonest sender does not know which part of the signature came from which participant, so any mismatches will be spread evenly between all participants.~Security against forging is guaranteed because all participants send their elements of the signature over secret channels, so no one, except the sender, can reproduce the full signature. 

The USS scheme proposed by Hanaoka \emph{et al.} \cite{HanaokaSig} makes some important improvements over \cite{ChaumUSS} at the cost of introducing a trusted authority who creates and distributes keys to each participant. {For~each of the $n$ users, $U_1, ..., U_n$, the trusted authority uniformly and randomly picks an $\omega$-vector, $v_i$, the components of which are elements of the finite field $F_q$. The trusted authority also constructs the~polynomial}
\begin{equation} \label{eq:poly}
F(x, y_1,...,y_\omega,z) = \sum^{n-1}_{i=0}\sum^\psi_{k=0}a_{i0k}x^iz^k+\sum^{n-1}_{i=0}\sum^\omega_{j=1}\sum^\psi_{k=0}a_{ijk}x^iy_jz^k
\end{equation}
{by choosing the coefficients $a_{ijk}$ uniformly at random from $F_q$.~The variables $x$, $y_i$ and $z$ are free variables in the polynomial, used for signing and verifying messages (see below).~It is assumed that both the message and each user's identity is described by an element in $F_q$, \emph{i.e.}, $m, U_i \in F_q$ for $i = 1,...,n$. The trusted authority then secretly distributes the following to each participant:}
\begin{enumerate}
\item [(1)]{A signing key: $s_i = F(U_i, y_1, ..., y_\omega,z)$};
\item[(2)] {A pair of verification keys: $v_i$ and $\tilde{v}_i = F(x, v_i,z)$}.
\end{enumerate}

Note that the signing keys, $s_i$, and the verification keys, $\tilde{v}_i$, are actually polynomials with coefficients in $F_q$. The identities of users (\emph{i.e.}, the $U_i\in F_q$) are public, while the signing and verification keys, as well as the polynomial $F$, are secret. To sign a message, $m$, user $U_i$ would send $(m, \alpha)$ to, say, user $U_j$, where $\alpha = F(U_i, y_1,...,y_\omega,m)$. To verify the message, user $U_j$ calculates
\begin{eqnarray*}
r_1 &=& F(U_i, v_j, m)\\
r_2 &=& \alpha |_{(y_1,...,y_\omega)=v_j}.
\end{eqnarray*}

{User $U_j$ accepts the message if and only if $r_1=r_2$. The security of this protocol derives from the fact that each participant has partial (but not full) knowledge of the polynomial $F$. Since their knowledge of the polynomial $F$ is limited, it is highly unlikely that they will be able to forge a message. Further, the participants do not know what knowledge other participants have, \emph{i.e.}, the verification keys are kept secret. It is this that guarantees transferability. In \cite{SwansonStinson}, a proof of the security of this scheme is presented in terms of their general security criteria.}

Unlike the original USS scheme, the scheme by Hanaoka \emph{et al.} in \cite{HanaokaSig} can be used to sign longer messages, and the length of the signature is shown to be optimal in the sense that it achieves a lower bound on the required memory size of a signature. A second advantage of this scheme is that it admits unlimited transferability between participants. This is in contrast to all of the quantum signature schemes presented below, as well as the USS in \cite{ChaumUSS}, where transferability requires the verification parameters $s_a, s_v$ to be such that $s_a < s_v$. Here, the parameter $s_a$ refers to a threshold used by a recipient who receives a message directly from a sender and $s_v$ to a threshold used for a forwarded message.

Given the existence of classical schemes that are unconditionally secure, one may wonder why quantum signatures are needed at all.~Although there is no definitive answer, so far it seems that quantum signature schemes are able to achieve the same functionality as classical USS, while making fewer assumptions. In any cryptographic protocol, assumptions are crucial to the practical viability and security of the scheme.~In \cite{ChaumUSS}, the resources assumed---an authenticated broadcast channel and secret authenticated classical channels---are expensive.~The secret authenticated channels between participants would require pairwise shared secret keys of the same length as the length of the messages being transmitted, {since information-theoretic secrecy requires application of the one-time pad}.~Further, it is known that between participants sharing only pairwise authenticated channels (even secret ones), a broadcast channel is only achievable if fewer than 1/3 of the participants are dishonest \cite{LamportByz}.~The improvements in \cite{HanaokaSig} come at the cost of introducing a trusted authority, whose role is to distribute the signing and verification keys to each participant. In reality, this makes the protocol vulnerable to targeted attacks against the trusted authority or even to dishonesty or incompetence on the part of the trusted authority. In contrast, the quantum signature protocols we will cover in this review do not assume either a broadcast channel or the existence of a trusted authority and are able to partly remove the need for secret classical channels by employing quantum channels instead.

We end this section by mentioning an ``almost classical'' signature scheme presented as ``P2''%please define
 in~\cite{DunjkoQKDComp}. It provides unconditional security while only requiring authenticated classical channels between all participants, as well as secret {classical} communication channels, which, for information-theoretic security, can be realised through QKD using untrusted, noisy quantum channels. Each pair of participants performs QKD using the noisy quantum channels and authenticated classical channels to generate shared secret keys, which can then be used to privately send classical messages {via the one-time pad}. For each future one-bit message, Alice will pick two uncorrelated random bit strings {of length $L$, which we denote as $A^0_B$, $A^1_B$, $A^0_C$ and $A^1_C$. The superscript denotes the future message, while the subscript denotes the participant to whom she will send the bit string. She will use the secret keys (generated using QKD) and the classical channels to privately send $A^0_B$, $A^1_B$ to Bob and $A^0_C$, $A^1_C$ to Charlie. Bob and Charlie will then use their secret keys (generated using QKD) to privately exchange half of their bit string elements with one another, thus symmetrising their keys from Alice's viewpoint. That is, Bob sends half of the bit values in $A^b_B$ to Charlie and receives half of the bit values in $A^b_C$ from Charlie, for $b=0,1$. To sign a message, $m$, Alice presents the message-signature pair $(m, A^m_B, A^m_C)$. The recipient, say Bob, will check that the signature matches the {parts} %halves 
of $A^m_B$, $A^m_C$ known to him. If there are fewer than $s_aL$ mismatches, he will accept the message-signature pair as valid. To forward a message, Bob will send $(b, A^m_B, A^m_C)$ to Charlie, who will perform the same checks, but will instead accept up to $s_vL$ errors, where $s_v>s_a$.}

Security against repudiation by Alice comes from the symmetrisation performed by Bob and Charlie. Since all communication is done classically, it is reasonable to set $s_a = 0$. In that case, for Bob to accept the message, he must find no mismatches with the {parts} %halves 
of $A^m_B$ and $A^m_C$ known to him. To successfully repudiate, Alice must also make Charlie receive more than $s_vL$ mismatches. Since Alice does not know which half of $A^m_B$, $A^m_C$ that Bob will keep/receive, any mismatch introduced has a $1/2$ chance of ending up with Bob and a $1/2$ chance of ending up with Charlie. To achieve this, an honest Bob/Charlie will not test for mismatches in the elements they forwarded. Simple probabilistic arguments show that
\begin{equation}
P(\text{Repudiation}) \leq (1/2)^{s_vL}.
\end{equation}

{In the case presented here, where Bob and Charlie exchange exactly half of their bit string elements, security against forging follows because Bob has zero information on the $L/2$ bits that Charlie received from Alice, but did not forward to Bob. In order to successfully forge a message, Bob must guess the values of these $L/2$ bits, making fewer than $s_vL$ mistakes.~The probability of Bob doing this can be bounded as}
\begin{equation}
P(\text{Forge}) \leq \exp(-(1/2 -s_v)^2L).
\end{equation}

The security of this protocol relies entirely on the security of the underlying QKD scheme used to generate the secret shared keys.~Secure QKD systems exist and are commercially available from a number of companies.~Therefore, this quantum signature protocol can be securely implemented using existing technology.~If the participants are using QKD to generate secret shared keys, they only need enough secret shared keys bits to authenticate their classical channels to initiate QKD at the start of the scheme.~QKD is then used for key expansion, to generate more shared secret keys.
``P2'' has been generalised to more than two recipients, which required the introduction of different levels of acceptance thresholds and separate checks for parts of the signatures received directly from Alice and for those forwarded from each other participant~\cite{JuanMiQCrypt}.

Note the similarities and differences between this protocol and most quantum signature protocols to be described below; in many respects, this is similar to other protocols described below, but in ``P2'', Alice sends {different} classical strings to Bob and Charlie, whereas in existing quantum signature protocols, Alice typically sends the same quantum states to all recipients.
``P2'' has the advantage of %significantly 
reducing the initial size of the secret shared keys as compared to quantum schemes, where Alice sends each recipient the same quantum states.

\section{Quantum Signature Schemes} \label{sec:quantum}

\subsection{Quantum One-Way Functions and Quantum Hash Functions}
\label{Sec:oneway}

Given how useful cryptographic hash functions are, it would be nice to find one that is provably a one-way function. In fact, quantum-mechanical unconditionally secure one-way functions exist, in the sense that given the description (e.g., wave function) of a quantum state, one can, at least in principle, prepare it. Given a copy of the quantum state, however, it is not even in principle possible to determine, with a 100\% success rate, exactly what that state is by measuring it (see, for example, the mapping given by Equation \eqref{eq:finger} below). More precisely, it is impossible to determine exactly what a quantum state is, unless one has very particular prior information about what the possible states are, namely if the only possible states are some set of orthogonal quantum states. Security is here guaranteed by the Holevo bound on the accessible information~\cite{Holevo, GottesmanQDS}.~Broadly speaking, this is what guarantees the security of all proposed quantum signature schemes.~(Note that a somewhat different definition of a quantum one-way function also has been made, as a function that is easily computable by a classical algorithm, but hard to invert even by a quantum computer~\cite{KashefiKerenidis}. Such a function does not necessarily have to involve quantum states.)

Apart from quantum signature schemes, beginning with the scheme in~\cite{GottesmanQDS}, the concept of quantum one-way or quantum hash functions has been used also for quantum fingerprinting~\cite{BuhrmanPrint, GavinskyPrint}. A fingerprint, loosely speaking, is a short identifier for a longer string. One can reformulate the properties of quantum one-way and quantum hash functions in the language of modern cryptography~\cite{Abla}, although this treatment does not emphasise that ``computing the quantum hash'' has to involve {preparation} of the quantum state, not just a computation of what the state should be, and, related to this, does not specify what is meant by ``inverting'' the quantum one-way or hash function. We therefore modify the definitions in~\cite{Abla} to read:

\begin{Definition}{Quantum one-way function:}
Let $\psi : \{0,1\}^n \rightarrow \mathcal{H}$ be the mapping $k \rightarrow |\psi_k\rangle$. Then, $\psi$ is called a quantum one-way function if it is easy to compute, \textit{i.e}., $|\psi_k\rangle$ for a particular $k$ can be determined using a polynomial-time algorithm, but impossible to invert, in the sense that if given a quantum state prepared in one of the states $|\psi_k\rangle$, one cannot, using any procedure allowed by quantum mechanics, with certainty determine which state one has been given.
\end{Definition}
{Security is %normally 
guaranteed by the Holevo bound~\cite{Holevo}. For example, if the quantum states have $m$ qubits and $k$ is an $n$-bit string where $n>m$, then it is impossible to obtain more than $m$ bits of information about $k$, {and depending on the states $|\psi_k\rangle$, the bound may be even tighter}. It is therefore impossible to perfectly determine $k$, and so it is also impossible to determine $|\psi_k\rangle$.}
\begin{Definition}{(n,s,$\delta$)-quantum hash function:}
Let $\psi$ be a quantum one-way function whose domain has size $2^n$ and whose range is a Hilbert space with dimension $2^s$. Suppose further that any distinct $w, w^\prime$ give $\delta$-orthogonal outputs ($\delta<1$), \textit{i.e}.,
$|\langle \psi(w) |\psi(w^\prime)\rangle| < \delta.$ Then, $\psi$ is an (n,s,$\delta$)-quantum hash function.
\end{Definition}

A classical-quantum or simply quantum hash function then is a function that satisfies all of the properties that a ``classical'' 
hash function should satisfy; such information-theoretically secure quantum one-way functions then exist.~Pre-image resistance follows from Holevo's theorem when $n>s$, since any measurement on $|\psi_k\rangle$ can reveal at most $s$ bits of information about $k$.~Second pre-image resistance and collision resistance follow, because all input states are mapped to states that are $\delta$-orthogonal. Therefore, we see that quantum hash functions can satisfy the three conditions with information-theoretic security. The above definitions are for discrete domains and finite-dimensional Hilbert spaces; we note that we can in principle also define quantum one-way and hash functions also for continuous variables, although this has, to our knowledge, not been formally done. Doing so may be useful, e.g., for outlining continuous-variable quantum signature schemes, in analogy to continuous-variable QKD.

A difference between classical and quantum hash functions arises in their collision properties.~For a classical hash function, a collision arises if two messages have the same hash.~For quantum hash functions, we may have {quantum collisions}, meaning that two quantum hashes (which are quantum states) might be different, but since non-orthogonal quantum states cannot be perfectly distinguished from each other, a test used to check whether quantum hashes are different may not always detect the difference.~It of course also matters what test we specify for testing the equality of quantum hashes; one possibility is quantum comparison~\cite{QuantCompSteve, QuantComp, AnderssonPRA2006}. {Another possibility is a SWAP 
 test, given in \cite{BuhrmanPrint}, which takes two states and an ancilla to perform the mapping}
\begin{equation}
|0\rangle |\phi\rangle |\psi\rangle \rightarrow \frac{1}{2}|0\rangle (|\phi\rangle |\psi\rangle + |\psi\rangle |\phi\rangle) + \frac{1}{2}|1\rangle (|\phi\rangle |\psi\rangle - |\psi\rangle |\phi\rangle).
\end{equation}

{Measurement of the ancilla qubit then produces outcome zero with certainty if the states are equal. The outcome one is obtained with probability $\frac{1}{2} - \frac{1}{2}|\langle\phi|\psi\rangle |^2$. If many copies of the two states are available, performing the SWAP test many times determines whether the states are equal or not with a probability that can be made arbitrarily close to one.}

As will be explained next, quantum one-way functions or cryptographic quantum hash functions can be used to generate unconditionally secure quantum signature protocols analogous to how cryptographic hash functions are used in the Lamport--Diffie one-time signature scheme.~Since quantum one-way functions are provably information-theoretically secure, the resulting quantum signature schemes are information-theoretically secure.

\subsection{Quantum Digital Signatures}

Gottesman and Chuang~\cite{GottesmanQDS} proposed the first quantum signature scheme, a ``quantum public key'' signature scheme in which the ``public keys'' are quantum states and the private keys are classical strings that specify what the quantum states are.~The scheme is an analogue of the classical Lamport--Diffie signature scheme, but with a quantum one-way function to generate unconditional security.~Quantum one-way and quantum hash functions have the additional property that the function output values (quantum hashes) can be made shorter than their classical counterparts, to give quantum ``fingerprints''~\cite{BuhrmanPrint}. 
To have unconditional security, it is actually not necessary for the function values to be short compared to the input strings. It is only necessary that the function values are non-orthogonal enough, so that an adversary cannot gain too much information about them. Nevertheless, short values are useful, since a scheme with shorter signatures is more practical.
If one defines the distance between two quantum states $|\psi\rangle$ and $|\psi'\rangle$ as $\sqrt{1-|\langle\psi|\psi'\rangle|^2}$, where $\langle\psi|\psi'\rangle$ is the inner product between the states, then there exist sets of states of $n$ qubits $\{|\psi_k\rangle\}$ satisfying $|\langle\psi_k|\psi_{k'}\rangle|\leq \delta$ for $k\neq k'$, such that the set may have many more than $2^n$ states if $\delta < 1$. Here, $k$ is the function input and $|\psi_k\rangle$ the corresponding output. Let us denote the length of the bit string $k$ by $L$; there are $2^L$ such bit strings. One possible such family of states are the quantum fingerprinting states suggested by Buhrman \emph{et al.}~\cite{BuhrmanPrint}, which gives $L=O(2^n)$ for $\delta \approx 0.9$. The resulting quantum one-way function is
\begin{equation} \label{eq:finger}
k \rightarrow |\psi_k\rangle = \frac{1}{\sqrt{m}} \sum^m_{i=1} (-1)^{E(k)_i} |i\rangle,
\end{equation}
where $E$ is a Justesen error-correcting code, mapping bit strings of length $L$ to bit strings of length $m$, such that $m = O(L)$, and $E(k)_i$ is the \emph{i}-th component of $E(k)$.~The right-hand side has dimension $m$ and can therefore be seen as a system of $\log_2(m) = n$ qubits.~In the terminology introduced above, the mapping is a $(L, n, \delta \approx 0.9)$-quantum hash function with $L = O(2^n)$.~Other possibilities mentioned in~\cite{GottesmanQDS} are quantum stabiliser states, with $L=n^2/2+o(n^2)$, or the single-qubit states $\cos (j\theta)|0\rangle + \sin (j\theta) |1\rangle$, for $\theta = \pi/2^L$ and integer $j$, for any $L$ and with $\delta = \cos \theta$.

To begin the protocol, the quantum public keys are first distributed to the recipients to enable the sender to later on send a signed message. {For the protocol to be secure, one must be careful about how the states are distributed. For clarity, this will be discussed below; for now, we note that the participants need a way of assuring that Alice has sent them identical quantum states and that the states have not been altered in transmission.}~When Alice wants to send a signed message $b$ to someone with access to the public key, $|\psi_k^b\rangle$, she will send the message along with $k^b$ (her signature). Given $k^b$, the receiver can apply the known quantum one-way function and compare the output to the public key. The message will be accepted if the public key matches the output of the one-way function. Just as in the Lamport scheme, at the end of the scheme, all used and unused quantum public and private keys must be discarded.

More concretely, to sign a single bit with $t$ participants, 
the protocol proceeds as follows:
\begin{enumerate}
\item[(1)] Alice chooses $M$ pairs of $L$-bit classical strings, $\{k^i_0,k^i_1\}$, $1\leq i\leq M$. The $k_0$'s will be used if the future message is chosen to be the bit zero and the $k_1$'s will be used if the future message is chosen to be the bit one. Increasing the value of $M$ will increase the security level of the protocol (security is exponential in $M$).
\item[(2)] Alice assigns each of the $L$-bit strings to a different element in the set of fingerprinting states, or whatever the chosen set of output quantum states is, according to a mapping known to all participants. That is, all participants know the one-way function, but not the $L$-bit strings used as input. She then 
distributes the quantum states to the $t$ participants, so that each participant has a suitable number of copies of each of the the $2M$ quantum states $\{ |\psi_{k^i_0}\rangle, |\psi_{k^i_1}\rangle\}$, $1\leq i\leq M$.
\item [(3)] Unless the distribution is managed by a trusted third party, the participants should perform some sort of test to ensure that they all received the same public keys.~{In the three-party setting,} Gottesman and Chuang suggest that {Alice sends two copies of each public key to each participant. Bob and Charlie would then both perform a SWAP test on their two keys to check that they are the same.~One participant, say Bob, would then pass one of his keys to Charlie, who would perform a SWAP test on this key and one of his own to determine if they are equal.~Bob performs a similar test on a key received directly from Alice and one forwarded by Charlie.~The keys used in these last SWAP tests would then be discarded, and Bob and Charlie are left with one copy of the public key each.~If any of the SWAP tests fail, the protocol is aborted.~If none of the tests fail, the participants have evidence that Alice did in fact send out the same public key states. By sending each participant more copies of the public key, the probability of discovering a cheating Alice can be made arbitrarily close to one.}

{In the more general $t$-party setting, the authors suggest a distributed symmetry test performed by all participants. In this case, Alice would distribute $t+1$ copies of the public key to each participant. They would perform a test to check for complete symmetry of their $t+1$ copies. If the test is passed, they would send one copy of the public key to each of the other $t-1$ participants, keeping one copy to use for signature verification and the remaining copy for a further symmetry test using all of the public keys received from other participants.}

\item[(4)] To later send the message $b$, Alice would send $(b, k^1_b, k^2_b, ..., k^M_b)$. From this, a recipient can easily compute $|\psi_{k^i_b}\rangle$ for each $i$ and compare the state to the public keys they previously received from Alice (again using a SWAP test). The recipient counts the number of mismatches he gets.
\item[(5)] If the mismatch rate is less than some rate $s_a$, the recipient will accept the message. If the message is forwarded on from another recipient, the mismatch rate must be less than $s_v$ to be accepted, where $s_v>s_a$.
\item[(6)] All used and unused keys are discarded.
\end{enumerate}

Security against forging comes from the fact that there are a limited number $T$ of copies of each quantum state in circulation, and each of the states has $n$ qubits. By Holevo's theorem, even if a forger gained access to all $T$ copies of the state $|f_{k^i_b}\rangle$, at best, he could discover $Tn$ bits of information about the $L$-bit classical string $k^i_b$ used as input to the quantum one-way function. If $Tn<L$, the forger will not be able to determine $k^i_b$ exactly. Furthermore, for unknown states, the SWAP test is only non-destructive if the states being compared are equal (the Holevo bound limits also this procedure). This means that the forger cannot apply a trial-and-error technique to discover $k^i_0$, since each time he is wrong, he will destroy one of his copies of the public key. Choosing the security parameters $s_a$ and $s_v$ to be low protects against forging attempts. 

Transferability comes from the fact that the public key needs to pass SWAP tests performed by the participants. Therefore, Alice can try to cheat by, for example, sending participants different quantum states, but unless what she sends is symmetric, it may not pass the SWAP tests.~The upshot of this is that Alice cannot engineer a state that will lead to a much higher average number of mismatches for one participant relative to any other participant. {If $M$ is large, the difference in the number of mismatches ({parts} of Alice's signature declaration that do not agree with the corresponding public key) observed by Bob and Charlie will be $O(\sqrt{M})$ with high probability, and it is therefore highly unlikely that one participant will observe fewer than $s_aM$ errors while the other observes more than $s_vM$ errors.} To ensure transferability, we need a sufficiently large gap between $s_v$ and $s_a$.

There are a few observations to make about this protocol.~First, a disadvantage of this and all other quantum signature schemes proposed so far is that the length of a signed message scales linearly with $L$, making it inefficient.~There seems to be no reason why this should be a fundamental limitation, though, and it is possible that more efficient protocols will emerge.~Second, as mentioned previously, this is a one-time signature scheme, meaning that the public/private key pair can only be used once. An interesting question is whether there is some method, possibly similar to the one used classically in~\cite{Merkle}, to enable reusability.~Lastly, the protocol as described is unfeasible with current technology.~This is because the distributed SWAP tests are non-trivial to perform, but above all, due to the requirement of long-term quantum memory.~The public keys are quantum states and must be kept indefinitely until Alice wants to send a message.~Methods of removing this requirement will be addressed in the next section.

We now further comment on the issue of how the public keys can be securely distributed. Of course, if a forger can make a recipient think that he has received a public key from Alice, when in fact it comes from the forger, then he can easily forge a message.~Therefore, it is clear that, as well as the SWAP or symmetry tests, there must be some sort of authentication of the quantum states being used as public keys. One solution proposed by Gottesman and Chuang is to use a trusted third party with authenticated links (as in \cite{BarnumAuth}) to all participants.~This simplifies the protocol by allowing the SWAP tests to be performed entirely by the third party, at the cost of introducing extra trust assumptions. Alternatively, Alice could directly send states to each recipient over authenticated quantum channels. Each solution requires authentication of quantum channels, which becomes very expensive for all practically implementable quantum digital signature schemes (discussed below). In Section \ref{sec:tamper}, we present an alternative key distribution method, which is much less expensive in terms of the resources consumed.

\subsection{Quantum Signatures Using a Multi-Port without Quantum Memory}

\begin{figure} \label{fig:Multiport}
\centering
\includegraphics[width=0.7\textwidth]{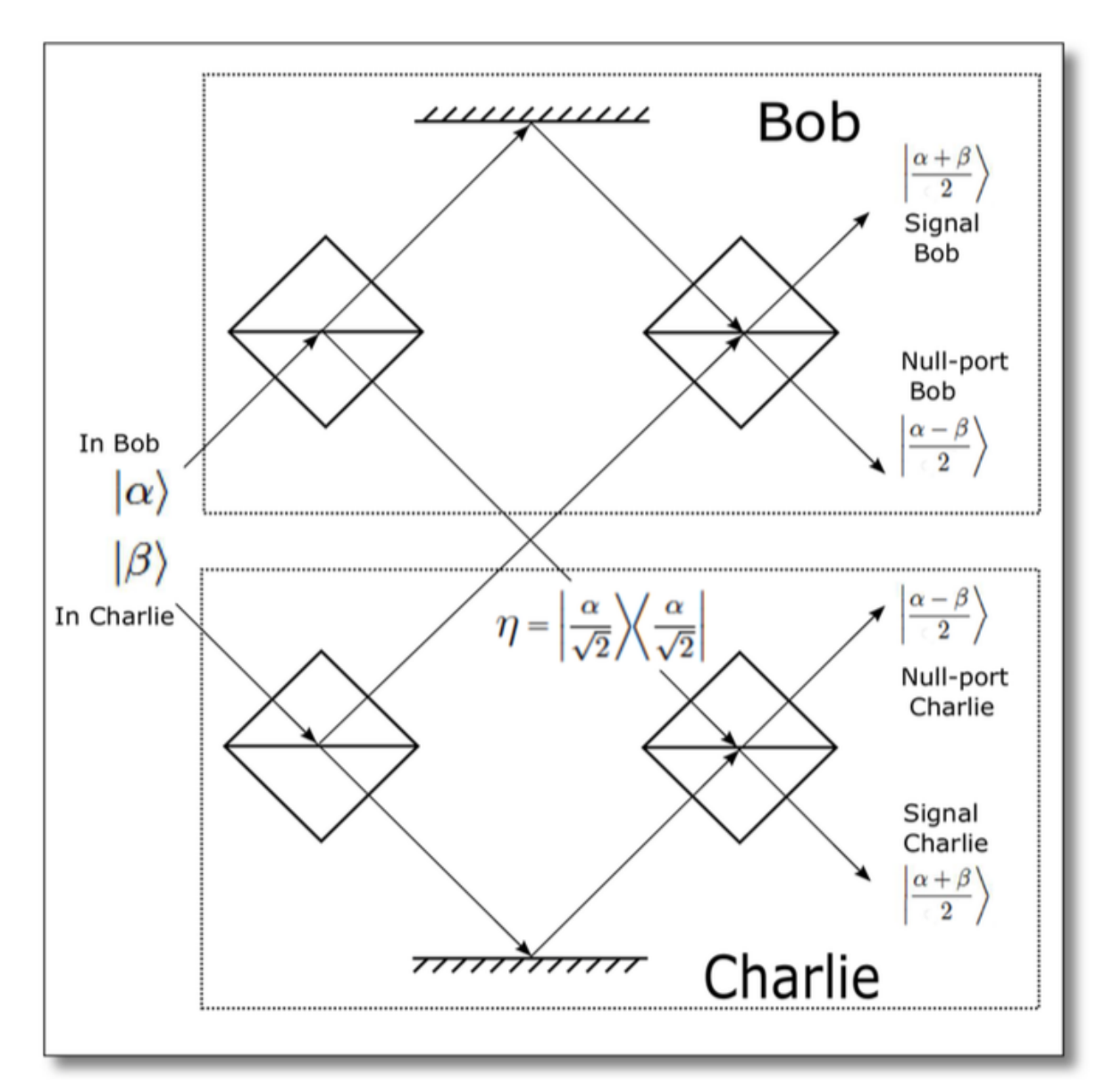}
\caption{Taken from \cite{DunjkoNoMem}. The multi-port used by Bob and Charlie to symmetrise the quantum states they receive from Alice.~Whatever overall quantum state Alice sends to Bob ($|\alpha\rangle$) and Charlie ($|\beta\rangle$), the outputs of the signal ports will be symmetric with respect to Bob and Charlie. Bob and Charlie can either store the signal port states in quantum memory or, alternatively, if one wishes to also remove the need for quantum memory, they can perform some type of quantum measurement on the signal ports directly when the states are received from Alice. Detecting photons in the null ports helps to detect cheating.}
\end{figure}

The original quantum signature protocol by Gottesman and Chuang~\cite{GottesmanQDS} has been further developed, in particular to make experimental realisation more feasible. In a scheme using coherent states outlined in \cite{AnderssonPRA2006}, message transferability is ensured by an optical multi-port, which is simpler to realise than a SWAP test.~The multi-port consists of four 50/50 beam splitters, as shown in Figure~\ref{fig:Multiport}.~It symmetrises Alice's input states, so that the recipients, here Bob and Charlie, will obtain the same measurement statistics. Alice therefore cannot make them disagree on the validity of a message. 
The symmetrisation procedure is non-destructive, that is, it does not alter the input state, if the total input quantum state is symmetric. This is the case in particular if the input states are identical. This allows the quantum states in the signal output modes to be further used; the null-ports then contain the vacuum state. If the input states are not identical (specifically, if the input state is not symmetric), then photons may be detected in the null-ports, and this can also be used to detect adverse activity, either by Alice or by a recipient.
This protocol was analysed and experimentally realised in \cite{CollinsNatComm}. The multi-port can in principle be generalised to more than two recipients, but as it essentially is an intertwined interferometer, the scheme becomes rather complex to implement.

An even more serious limitation, however, was the requirement of long-term quantum memory. The scheme proposed in \cite{DunjkoNoMem} and experimentally realised in \cite{CollinsPRL} removes this requirement.~Here, the recipients immediately measure the quantum states they receive from Alice and only store the classical outcomes.~This way, their test that the declared private key matches the previously received quantum public key is a little less efficient at detecting mismatches than if they keep their quantum states in memory until they know what state they are supposed to test for, but this is a small price to pay for removing the need for quantum memory. Importantly, security is still exponential in the length of the sequence of states that Alice selects.

In all of these schemes, just as in the scheme by Gottesman and Chuang, the quantum signature states are first distributed, enabling signed messages to be sent later on.~The distribution stage typically takes place long before the messaging stage, perhaps weeks or months.~The schemes in \cite{AnderssonPRA2006, DunjkoNoMem} are phrased in terms of coherent states, 
although any set of non-orthogonal quantum states could in principle be used.
If using coherent states, in order to be able to later sign a one-bit message sent to Bob or Charlie according to the scheme in \cite{DunjkoNoMem}, Alice would choose two $L$-bit strings, $k_0$ and $k_1$, one for each future message. She encodes the zeros in this bit string as the low-intensity coherent state $|\alpha\rangle$ and the ones as $|-\alpha\rangle$. That is, the bit strings Alice chooses determine the sequences of phases for the coherent states. Alice's signature for message $b$ is $k_b$, and these bit strings must be kept secret. Alice could also choose from more than two different phases; in the corresponding experimental realisation in \cite{CollinsPRL}, she chooses from four different phases. Alice sends one copy of the sequence of coherent states to Bob and one to Charlie. As in~\cite{AnderssonPRA2006, CollinsNatComm} and also in \cite{DunjkoNoMem}, Bob and Charlie pass the states through the same type of optical multi-port in order to symmetrise the states that they receive, thus ensuring the transferability of messages.~Directly when receiving the states, however, they both measure the states and record the classical outcomes. The classical outcomes are Bob's and Charlie's keys, which they will keep secret and use to verify Alice's signature, testing for mismatches, much as in previous schemes.~In this way, they each have partial information about Alice's signature. Exactly what information Bob and Charlie have about Alice's signatures after the distribution stage depends on what measurement they use. In~\cite{DunjkoNoMem}, unambiguous quantum state discrimination (USD)~\cite{SteveBook} is suggested.~When this type of quantum measurement gives a result, it is guaranteed to be correct, but sometimes, the measurement fails to give any result.~Then, depending on the outcome of the USD measurement, they each independently gain either full information on Alice's quantum state or zero information.

When Alice wants to send a message $b$ to Bob, she will send $(b, k_b)$. Just as before, Bob will check $k_b$ against his measurement outcomes and only accept the message if there are fewer than $s_aL$ mismatches, with some suitably chosen $s_a$. To forward the message to Charlie, he sends exactly what he received from Alice. Charlie performs the same tests, but uses the mismatch threshold $s_vL$ instead. Just as before, choosing a low value of $s_v$ gives protection against forging attempts, {since in order to forge a message, Bob would have to guess a message-signature pair that has fewer than $s_vL$ mismatches with Charlie's measurement outcomes. It is shown in \cite{DunjkoNoMem} that for individual or collective attacks, the probability of a recipient, say Bob, being able to find a message signature pair (not originating from Alice) that Charlie will accept as valid can be bounded by}
\begin{equation}
P(\text{Forge}) \leq \exp \left( -2\left( p_{min} - s_v \frac{p_{USD}}{p_{USD}-\delta } \right)^2 (p_{USD}-\delta)L\right),
\end{equation}
{where $p_{min}$ is the minimum error probability, \emph{i.e.}, the minimum probability achievable that Bob incorrectly identifies a quantum signature element sent to him by Alice. The probability of obtaining an unambiguous outcome in the USD measurement is $p_{USD}$, and $\delta$ is a small parameter that takes into account the worst-case scenario, whereby Charlie's unambiguous state discrimination measurement is successful only $(p_{USD}-\delta)L$ times (any fewer, and the protocol would be aborted).~The $p_{min}$ term arises because Bob's optimal strategy would be to perform a minimum-error measurement on each of the states he received from Alice in order to make the best possible guess of the states Alice actually sent.} 

Choosing a large gap between $s_v$ and $s_a$ gives protection against repudiation and provides transferability. {It is shown in \cite{DunjkoNoMem} that even for the most general cheating strategies available to Alice, the probability of successful repudiation decays exponentially in the size of the signature length. More specifically, it can be shown that the probability of Alice being able to send a sequence of states such that one recipient receives fewer than $s_aL$ mismatches while the other receives more than $s_vL$ mismatches with her signature declaration is bounded as}
\begin{equation}
P(\text{Repudiation}) \leq \exp\left(-\frac{1}{2}p^2_{USD}(s_v-s_a)^2L\right).
\end{equation}

{This is because the multi-port ensures that the overall state sent by Alice is symmetric with respect to the exchange of Bob and Charlie. Therefore, any mismatches introduced by Alice are equally likely to be discovered by Bob or Charlie, and so, the probability of one finding less than $s_aL$ errors and the other finding more than $s_vL$ errors becomes small. If there are no imperfections, one can choose $s_a=0$, but as soon as the implementation is not ideal, one must choose $s_a>0$; otherwise, even an honest Alice may not be able to sign messages. Note that finding optimal choices of the parameters $s_a$ and $s_v$ is highly non-trivial and depends on many practical aspects of the experimental setup.}

In the experimental realisation, Bob and Charlie instead use unambiguous quantum state elimination, which is a quantum measurement that unambiguously rules out one or more of the states that Alice has chosen~\cite{SteveBook, statelim, CollinsPRL}. This measurement has a higher success rate than USD, making the signature scheme more efficient. Importantly, no matter what type of measurement Bob and Charlie use, neither of them knows Alice's full signature, nor do they know exactly what the other recipient knows about Alice's signature. Furthermore, Alice does not know what Bob knows and what Charlie knows. Loosely speaking, this guarantees security against forging and repudiation, although security against coherent forging attacks to date remains formally unproven for schemes using coherent states. In coherent attacks, a forger can make measurements in an entangled basis on any number of the quantum states that Alice sends. In individual and collective attacks, a forger is limited to measurements on individual states or classically correlated measurements on individual states.

The protocol removing the need for quantum memory was an important step towards implementable quantum digital signatures. However, the experimental demonstration \cite{CollinsPRL} showed that efficient implementation is difficult; for $\alpha=1$, in order to sign a single ``half-bit'' with a security level of $99.99\%$, a value of $L = 5.10\times10^{13}$ was required. This was mainly attributed to the multi-port causing large losses and being difficult to align.

\subsection{Quantum Signatures with Quantum Key Distribution Components}

In Wallden \textit{et al}.~\cite{DunjkoQKDComp}, to further make quantum signatures more feasible,
the multi-port is removed. Furthermore, the protocol is phrased in terms of BB84 states instead of coherent states. The BB84 states are more convenient to work with, as security proof techniques can be leveraged from work on relativistic quantum bit commitment~\cite{AdrianBit}. This enabled the first proof of security against coherent forging.

Apart from the absence of the multi-port, the protocol works similarly to before.~For each future possible message, Alice will create two copies of a classical string of symbols drawn from the set $\{0,1,+,-\}$. This will be her signature, and she must keep it secret. She encodes each symbol into the BB84 quantum states $\{|0\rangle, |1\rangle,|+\rangle,|-\rangle\}$, where $|\pm\rangle = (|0\rangle \pm |1\rangle)/\sqrt{2}$, and sends one copy of the sequence of states to Bob and one to Charlie. The measurements made by Bob and Charlie are quantum unambiguous state elimination (USE) measurements.
Each of them can with certainty rule out one of the possible states for each position in the sequence. For example, if Bob makes a measurement in the $Z$ basis and obtains the result $|0\rangle$, then he knows that Alice cannot have sent the state $|1\rangle$, but could have sent any of the other three states. Without knowing the encoding basis for a state, Bob and Charlie cannot make a measurement that distinguishes exactly which state Alice sent. In this way, Bob and Charlie each gain partial information of Alice's signature, also without knowing exactly what information the other recipient has gained. Note that in contrast to QKD, the parties do not proceed to announce their measurement and preparation bases.

Previously, the purpose of the multi-port was to symmetrise the states Alice sent to Bob and Charlie in order to protect against repudiation. In fact, this symmetrisation can be achieved just as well by requiring that with probability 1/2, rather than measuring a state received from Alice, Bob/Charlie would instead 
forward it to the other participant. In this way, Bob receives about half the states that Alice sent to him and about half the states that Alice sent to Charlie and \textit{vice versa}. Importantly, it is assumed that Alice cannot eavesdrop on the Bob-Charlie quantum channel, since she should not know who has which state. This means that from her point of view, the overall state is symmetric with respect to exchange of Bob and Charlie, which provides security against repudiation similar to before. Alternatively, Bob and Charlie can measure all quantum states they receive from Alice and then use a classical secret communication channel to randomly forward some results to the other recipient.

Mainly because the multi-port is removed, this protocol achieves a significant improvement in efficiency over the previous one.~In the latest experimental realisation \cite{CollinsNoMulti}, for the same security requirements as in \cite{CollinsPRL}, the length of the required signature is reduced to $L = O(10^{9})$.~Further, as mentioned above, the protocol is secure even against coherent forging. Still, further improvements in efficiency are needed, and signature length still scales linearly with message length. More efficient protocols would be desirable. Furthermore, as we will discuss next, an important security assumption would need to be relaxed.

\subsection{Security against Tampering with the Quantum Channels} \label{sec:tamper}

All quantum signature protocols described so far have made a strong assumption that we have not mentioned until now:~that all quantum channels are ``tamperproof'', with the expectation that this assumption could be removed if the parties use a procedure similar to parameter estimation in QKD~\cite{QKDreview, QKDreview2}.~By ``tamperproof'', we mean that it is guaranteed that the participants do not eavesdrop or otherwise tamper with the quantum channels used by Alice to send quantum states to Bob and Charlie and, if applicable, with the channels used by Bob to forward quantum states to Charlie and \textit{vice versa}. In parameter estimation for QKD, the parties declare some of the transmitted states and obtained measurement results, to check for errors that an eavesdropper would cause. This allows them to bound the information an eavesdropper could hold about the remaining undeclared states and the resulting key.

In \cite{AmiriQSig}, this strong assumption on the quantum channels is removed. {The protocol requires insecure quantum channels connecting all participants pairwise, as well as pairwise authenticated classical channels.~Alice-Bob and Alice-Charlie separately perform the BB84 QKD protocol, but without the classical post-processing steps of error correction and privacy amplification. We call this sub-protocol the key generation protocol (KGP), and {shared keys are generated} for each possible future message. Following the Alice-Bob KGP, Alice has bit strings $A^0_B$ and $A^1_B$, where the superscript denotes what the future message bit string is for, and the subscript denotes that the KGP was performed with Bob. Bob holds the strings $B^0$ and $B^1$, {which ideally would be identical to $A^0_B$ and $A^1_B$, but in general, will not be}. Parameter estimation can be used to estimate the correlation between Alice and Bob's bit strings, and from this, the level of possible eavesdropping can be quantified just as in QKD. As long as the error rate between Alice's and Bob's string is sufficiently low, it can be shown that, in the case of individual and collective attacks, any potential eavesdropper cannot (except with negligible probability) produce a string that is more correlated with $B^b$ than $A^b_B$ is correlated with $B^b$, for $b=0,1$. Exactly the same arguments apply to the Alice-Charlie KGP, denoting Charlie's bit strings by $C^b$ for $b=0,1$.}

{As in previous protocols, Bob and Charlie symmetrise their bit strings to ensure security against repudiation. That is, Bob privately sends half of $B^b$ to Charlie and privately receives half of $C^b$ for $b=0,1$. To send a message, $m$, Alice would send $(m, A^m_B, A^m_C)$ to the desired recipient. The recipients then verify the signature as before, using verification thresholds $s_a$ and $s_v$. A detailed {description of the} security analysis is beyond the scope of this review, but is provided in \cite{AmiriQSig}.}

{A major difference in this protocol compared to previous protocols is that Alice no longer sends the same sequences of states to Bob and Charlie, but instead sends them different sequences. Importantly,} Bob and Charlie can still guard against repudiation by randomly exchanging part of their measurement results, secret from Alice. However, the advantage is that a forger no longer has full access to a legitimate copy of Alice's quantum sequence that she sent to the other participant. This means that the protocol will require shorter state sequences than previous protocols where Alice sent all recipients the same quantum states. The ``partial QKD'' procedure also means that this quantum signature scheme in practice can be performed over quantum channels, which are too imperfect for QKD to be possible (although the Bob-Charlie quantum channel {used for the symmetrising exchange} must still be good enough to perform QKD). This is because in a practical implementation of QKD, error correction will somewhat decrease the threshold for the acceptable quantum bit error rate. For a signature scheme, it is not essential to distil an error-free and perfectly secret shared key. It is instead enough that Alice's declaration of her signature gives rise to fewer mismatches than a declaration made by a forger.

This protocol is no longer a ``quantum public key'' signature protocol, since all recipients now receive different ``quantum keys''.~Nevertheless, compared with the previous ``quantum public key'' signature protocols, this seems to be a minor disadvantage, as the signature distribution stage is no more complicated than when Alice sends every recipient the same quantum state sequences. 

\subsection{Coherent State Mappings}

An interesting connection between the quantum signature protocols using coherent states~\cite{CollinsNatComm, DunjkoNoMem} and the original Gottesman--Chuang protocol~\cite{GottesmanQDS} is highlighted in \cite{ArrazolaComm}. The paper uses techniques from \cite{ArrazolaFinger} to map the fingerprinting states in Equation~\eqref{eq:finger} to trains of coherent states. Written in the form given in Equation~\eqref{eq:finger}, the fingerprinting states are viewed as living in an $m$-dimensional Hilbert space. Assuming $m$ is a power of two, we can decompose this large Hilbert space into a tensor product of two-dimensional Hilbert spaces to get an equivalent representation, expressed as $\log_2(m)$ qubits. However, the high degree of entanglement between the qubits makes such a state very difficult (or impossible) to create with current technology. Alternatively, the state could be viewed as arising from a single photon distributed over $m$ orthogonal optical modes. In this case, the fingerprinting states can be exactly expressed as
\begin{equation} \label{eq:single}
\frac{1}{\sqrt{m}} \sum^m_{i=1}(-1)^{E(k)_i} |1\rangle_i,
\end{equation}
where $|1\rangle_i$ represents one photon in the \emph{i}-th mode. Finally, instead of considering a single photon, the authors suggest considering the train of coherent states
\begin{equation}
|\alpha\rangle_k = \bigotimes^m_{i=1} \left| (-1)^{E(k)_i}\frac{\alpha}{\sqrt{m}}\right\rangle_i,
\end{equation}
where $\alpha$, the coherent state amplitude, will be chosen to be small. When this state is projected onto the single-photon subspace, the single-photon Expression \eqref{eq:single} is recovered. 

The advantage of this mapping to trains of coherent states is that it gives a simple and experimentally practical method to create and compare the fingerprinting states for quantum signatures.~This comes at the cost of having a large number, $m$, of modes.~Further, it reveals a close similarity between the Gottesman--Chuang protocol and the coherent-state multi-port protocols defined above.~In the Gottesman--Chuang protocol, Alice may choose a fingerprinting state to act as the quantum public key and sends a copy to both Bob and Charlie.~Bob and Charlie use a SWAP test to ensure that they received the same state and then store the quantum public key.~If Alice were to apply the coherent state mapping to the quantum public key, she would then send a train of coherent states to both Bob and Charlie. The equality of the coherent states can be tested using the multi-port, analogous to the SWAP test. Alternatively, some sort of parameter estimation procedure could be used. Finally, rather than storing the states, Bob and Charlie could measure the states and store the classical outcomes. In this way, they each gain partial information on the quantum public key, but not full information, and so, cannot forge. The amount of information they gain will depend on the value of $\alpha$.

\section{Conclusions}

Signature schemes are widely used in modern communications.~Despite their importance, research into unconditionally secure schemes has remained a niche field in quantum cryptography and modern ``classical'' cryptography alike.~This is largely due to the highly significant advantages, both in terms of efficiency and ease of use, that currently used public key signature protocols have over any unconditionally secure schemes.~Nevertheless, with the advent of quantum computers, or better algorithms, RSA, DSA and ECDSA would all become obsolete, and signatures would have to be generated in other ways. At this point, hash-based signature schemes are a possible option. Schemes of this type may also prove to be resistant to attacks by quantum computers, depending on the one-way function used. However, since their security rests on the difficulty of inverting certain functions, these schemes are still only computationally secure, 
and there is no fundamental reason why future advances will not render them insecure.

In some highly sensitive applications, it may be desirable to have unconditional security instead. Research suggests that quantum mechanical features can be used to construct unconditionally secure signature schemes, such as the ones covered in this short review paper, that seem to require fewer additional resources than their classical counterparts. So far, this has come at the cost of lower efficiency, though this may be because quantum schemes are still largely unexplored. While the classical protocol ``P2'' in \cite{DunjkoNoMem} at the moment seems to be the most complete and easily implementable quantum signature protocol presented, it may not be the most efficient. For example, the Alice-Bob QKD link generates a perfectly secret shared key, when in fact, all that is required is that any eavesdropper is less correlated with Bob's key than Alice is. As mentioned above, it can indeed be shown \cite{AmiriQSig} that quantum signatures are possible with quantum channels that are of too low quality for QKD to be practical.

As it stands, in all quantum signature schemes, the signature length scales linearly with the size of the message.~Efficiency would be greatly increased if this scaling could be improved, and in fact, our recent investigations seem to suggest a scheme whose signature length 
scales more favourably with the size of the message being transmitted. The length of the signature is not the only measure of efficiency though: other respects in which protocols may be optimised are in the number of quantum channels required between participants and the number of secret pre-shared bits required between participants. Currently, all quantum schemes require pairwise quantum channels between all participants. When the number of participants becomes high, this requirement becomes expensive, and it would be interesting to see if a practical protocol exists requiring fewer quantum channels.

\acknowledgments
The authors acknowledge helpful discussions with Petros Wallden. Ryan Amiri acknowledges support by the EPSRC CM-DTC scheme, and Erika Andersson acknowledges partial support by EPSRC EP/K022717/1 and EP/M013472/1.

\end{document}